%% file: main.tex
\documentclass[]{spie}  %>>> use for US letter paper
\usepackage{amsmath,amsfonts,amssymb}
\usepackage{graphicx}
\usepackage[colorlinks=true, allcolors=blue]{hyperref}
\usepackage{subfig}

\input{defs}

\title{Analysis of Temperature-to-Polarization Leakage in BICEP3 and Keck CMB Data from 2016 to 2018}

\begin{document} 

\input{./authors}

\maketitle

\input{./abstract}
\input{./sec_intro}
\input{./sec_optical}
\input{./sec_ffbm}
\input{./sec_beamsim}
\input{./sec_conclusion}

\acknowledgments % equivalent to \section*{ACKNOWLEDGMENTS}       
 
The BICEP/\textit{Keck} project (including BICEP2, BICEP3 and BICEP Array) have been made possible through a series of grants from the National Science Foundation including 0742818, 0742592, 1044978, 1110087, 1145172, 1145143, 1145248, 1639040, 1638957, 1638978, 1638970, 1726917, 1313010, 1313062, 1313158, 1313287, 0960243, 1836010, 1056465, \& 1255358 and by the Keck Foundation. The development of antenna-coupled detector technology was supported by the JPL Research and Technology Development Fund and NASA Grants 06-ARPA206-0040, 10-SAT10-0017, 12-SAT12-0031, 14-SAT14-0009, 16-SAT16-0002, \& 18-SAT18-0017. The development and testing of focal planes were supported by the Gordon and Betty Moore Foundation at Caltech. Readout electronics were supported by a Canada Foundation for Innovation grant to UBC. The computations in this paper were run on the Odyssey cluster supported by the FAS Science Division Research Computing Group at Harvard University. The analysis effort at Stanford and SLAC was partially supported by the Department of Energy, Contract DE-AC02-76SF00515. We thank the staff of the U.S. Antarctic Program and in particular the South Pole Station without whose help this research would not have been possible. Tireless administrative support was provided by Kathy Deniston, Sheri Stoll, Irene Coyle, Donna Hernandez, and Dana Volponi.

% References
\bibliographystyle{spiebib} % makes bibtex use spiebib.bst
\bibliography{report} % bibliography data in report.bib

\end{document}

%% file: defs.tex
\def\bicep{{\sc Bicep}}

\def\biceptwo{{\sc Bicep2}}
\def\bicepthree{{\sc Bicep3}}
\def\bk{{\sc Bicep/\it Keck Array}}
\def\keck{{\it Keck Array}}
\def\biceparray{{\sc Bicep} Array}

\def\deg{^\circ}
\def\tp{$T \rightarrow P$}

%% file: authors.tex
% List of authors
% Wait for final decision on list from PIs

\author[a,b]{T. St. Germaine}
\affil[a]{Center for Astrophysics $\vert$ Harvard \& Smithsonian, Cambridge, MA 02138, U.S.A}
\affil[b]{Department of Physics, Harvard University, Cambridge, MA 02138, USA}
\author[c]{P.~A.~R.~Ade}
\affil[c]{School of Physics and Astronomy, Cardiff University, Cardiff, CF24 3AA, United Kingdom}
\author[d,e]{Z.~Ahmed}
\affil[d]{SLAC National Accelerator Laboratory, 2575 Sand Hill Road, Menlo Park, CA 94025}
\affil[e]{Kavli Institute for Particle Astrophysics and Cosmology, Stanford University, 452 Lomita Mall, Stanford, CA 94305}
\author[f]{M.~Amiri}
\affil[f]{Department of Physics and Astronomy, University of British Columbia, Vancouver, British Columbia, V6T 1Z1, Canada}
\author[a,g]{D.~Barkats}
\affil[g]{Institut Laue-Langevin, 38042 Grenoble Cedex 9, France}
\author[h]{R.~Basu Thakur}
\affil[h]{Department of Physics, California Institute of Technology, Pasadena, California 91125, USA}
\author[i]{C.~A.~Bischoff}
\affil[i]{Department of Physics, University of Cincinnati, Cincinnati, Ohio 45221, USA}
\author[h,j]{J.~J.~Bock}
\affil[j]{Jet Propulsion Laboratory, Pasadena, California 91109, USA}
\author[a]{H.~Boenish}
\author[k]{E.~Bullock}
\affil[k]{Minnesota Institute for Astrophysics, University of Minnesota, Minneapolis, 55455, USA}
\author[l]{V.~Buza}
\affil[l]{Kavli Institute for Cosmological Physics, University of Chicago, Chicago, IL 60637, USA}
\author[m]{J.~R.~Cheshire}
\affil[m]{School of Physics and Astronomy, University of Minnesota, Minneapolis, 55455, USA}
\author[n]{J.~Connors}
\affil[n]{National Institute of Standards and Technology, Boulder, Colorado 80305, USA}
\author[a]{J.~Cornelison}
\author[m]{M.~Crumrine}
\author[e,d,o]{A.~Cukierman}
\affil[o]{Department of Physics, Stanford University, Stanford, California 94305, USA}
\author[n]{E.~Denison}
\author[a]{M.~Dierickx}
\author[p]{L.~Duband}
\affil[p]{Service des Basses Temp\'{e}ratures, Commissariat \`{a} lEnergie Atomique, 38054 Grenoble, France}
\author[a]{M.~Eiben}
\author[f]{S.~Fatigoni}
\author[q,r]{J.~P.~Filippini}
\affil[q]{Department of Physics, University of Illinois at Urbana-Champaign, Urbana, Illinois 61801}
\affil[r]{Department of Astronomy, University of Illinois at Urbana-Champaign, Urbana, Illinois 61801, USA}
\author[m]{S.~Fliescher}
\author[e,o]{N.~Goeckner-Wald}
\author[a]{D.~C.~Goldfinger}
\author[o]{J.~A.~Grayson}
\author[a]{P.~Grimes}
\author[m]{G.~Hall}
\author[f]{M.~Halpern}
\author[a]{S.~A.~Harrison}
\author[d,e]{S.~Henderson}
\author[h,j]{S.~R.~Hildebrandt}
\author[n]{G.~C.~Hilton}
\author[n]{J.~Hubmayr}
\author[h]{H.~Hui}
\author[d,e,o,n]{K.~D.~Irwin}
\author[h,o]{J.~Kang}
\author[l]{K.~S.~Karkare}
\author[o]{E.~Karpel}
\author[h]{S.~Kefeli}
\author[o]{S.~A.~Kernasovskiy}
\author[a,b]{J.~M.~Kovac}
\author[o,d,e]{C.~L.~Kuo}
\author[m]{K.~Lau}
\author[l]{E.~M.~Leitch}
\author[j]{K.~G.~Megerian}
\author[h]{L.~Minutolo}
\author[h]{L.~Moncelsi}
\author[m]{Y.~Nakato}
\author[s]{T.~Namikawa}
\affil[s]{Department of Applied Mathematics and Theoretical Physics, University of Cambridge, Cambridge CB3 0WA, UK}
\author[h,j]{H.~T.~Nguyen}
\author[h,j]{R.~O'Brient}
\author[o]{R.~W.~Ogburn~IV}
\author[i]{S.~Palladino}
\author[m]{N.~Precup}
\author[p]{T.~Prouve}
\author[m,k]{C.~Pryke}
\author[a]{B.~Racine}
\author[n]{C.~D.~Reintsema}
\author[a]{S.~Richter}
\author[h]{A.~Schillaci}
\author[a]{B.~L.~Schmitt}
\author[m]{R.~Schwarz}
\author[k]{C.~D.~Sheehy}
\author[h]{A.~Soliman}
\author[h]{B.~Steinbach}
\author[c]{R.~V.~Sudiwala}
\author[t]{G.~P.~Teply}
\affil[t]{Department of Physics, University of California at San Diego, La Jolla, California 92093, USA}
\author[e,o]{K.~L.~Thompson}
\author[o]{J.~E.~Tolan}
\author[c]{C.~Tucker}
\author[j]{A.~D.~Turner}
\author[q,r]{C.~Umilt\`{a}}
\author[l]{A.~G.~Vieregg}
\author[h]{A.~Wandui}
\author[j]{A.~C.~Weber}
\author[f]{D.~V.~Wiebe}
\author[m]{J.~Willmert}
\author[a,b]{C.~L.~Wong}
\author[d,e,o]{W.~L.~K.~Wu}
\author[o]{E.~Yang}
\author[o,d,e]{K.~W.~Yoon}
\author[e,d,o]{E.~Young}
\author[o]{C.~Yu}
\author[a]{L.~Zeng}
\author[h]{C.~Zhang}
\author[h]{S.~Zhang}

\authorinfo{Send correspondence to T St. Germaine: stgermaine@g.harvard.edu}

%% file: abstract.tex
\begin{abstract}

The \bk\ experiment is a series of small-aperture refracting telescopes observing degree-scale Cosmic Microwave Background polarization from the South Pole in search of a primordial $B$-mode signature.  
As a pair differencing experiment, an important systematic that must be controlled is the differential beam response between the co-located, orthogonally polarized detectors.  
We use high-fidelity, \textit{in-situ} measurements of the beam response to estimate the temperature-to-polarization (\tp) leakage in our latest data including observations from 2016 through 2018.  
This includes three years of \bicepthree\ observing at 95 GHz, and multifrequency data from \keck. 
Here we present band-averaged far-field beam maps, differential beam mismatch, and residual beam power (after filtering out the leading difference modes via deprojection) for these receivers.
We show preliminary results of ``beam map simulations,” which use these beam maps to observe a simulated temperature (no $Q/U$) sky to estimate \tp\ leakage in our real data.

\end{abstract}

\keywords{Inflation, Gravitational waves, Cosmic microwave background, Polarization, BICEP, Keck Array}

%% file: sec_intro.tex
\section{INTRODUCTION}
\label{sec:intro}  

Observations of the Cosmic Microwave Background (CMB) over the past few decades have given us a key view of the early universe.  
These observations have played a central role in the development and validation of the $\Lambda$CDM standard model of cosmology.
Since the discovery of the 2.7 K CMB  \cite{Penzias65} and its 3 mK dipole \cite{Conklin69}, there have been detections of $\sim$ 100 $\mu$K temperature anisotropies \cite{Smoot92}, $\sim$ 1 $\mu$K $E$-mode anisotropies \cite{Kovac02}, and $\sim$ 100 nK $B$-mode anisotropies due to gravitational lensing of $E$ modes \cite{Hanson13}.  
There may exist a fainter $B$-mode signal in the polarization of the CMB due to the presence of gravitational waves at the time of recombination.  
The resulting $B$-mode signal peaks at degree angular scales, and is parametrized by the tensor-to-scalar ratio $r$. 
A detection of the $B$-mode signal due to primordial gravitational waves would constitute strong evidence for the existence of this period of inflation.

The \bk\ (BK) series of CMB polarization experiments has been observing the sky in search for this signal from the Amundsen-Scott South Pole Station since 2006 \cite{Barkats14}.  
We use small-aperture, on-axis refracting telescopes to search for the primordial $B$-mode signal at its $\ell \sim$ 100 peak.
The compact telescope design also allows us to observe the CMB at multiple boresight rotation angles, which provides valuable cross-checks and systematics control.
The optical elements are cryogenically cooled to 4 K / 50 K to minimize internal loading (see Section \ref{sec:optical} for more details on optical design).
The detectors are antenna-coupled transition-edge sensor (TES) bolometers that are cooled to 250 mK using a three-stage Helium sorption fridge.
With each new telescope in our program, we have increased the detector count in order to reduce the statistical uncertainty and increase mapping speed.
\bicep2\ held 512 detectors at 150 GHz; \keck\ featured five receivers each housing 512 (150/220/270 GHz) or 288 (95 GHz) detectors; \bicep3, with its larger aperture, holds 2,560 detectors at 95 GHz.
Prior to the start of the 2020 observing season, we installed the first  \biceparray\ receiver, which includes 192 detectors at 30 GHz and 300 detectors at 40 GHz \cite{spie:hhui}.  
Three other \biceparray\ receivers at 95, 150, and 220/270 GHz are currently in development.
Our most recent published results, which includes observations from \biceptwo\ and \keck\ at 150, 95, and 220 GHz through 2015 (BK15), yields a constraint of $r < 0.06$ when combined with \textit{Planck} temperature measurements\cite{BK15}. 

At the focal plane, optical power couples to two co-located, orthogonally polarized planar antenna arrays, passes through an on-chip band-defining filter, and is eventually dissipated on a suspended bolometer island and detected by two TES bolometers.  
This defines a single detector pair.  
Any mismatch in the beam pattern between the two polarization states (which we call $V$ and $H$ in this paper) in a detector pair leaks temperature anisotropies to the polarization measurement.
This prominent systematic, temperature-to-polarization (\tp) leakage, must be accounted for, as it may bias the final $r$ estimate.  
Most of the leakage power, coming from the lowest-order main beam difference modes, is filtered out in analysis using a technique called deprojection (discussed in Section \ref{sec:ffbm}).
We also utilize our vast library of far-field beam measurements in specialized ``beam map simulations," which estimates the higher-order, undeprojected \tp\ leakage in the CMB data.
In the BK15 data set, when the measured \tp\ leakage is added to simulations, the resulting bias on $r$ recovered from multicomponent likelihood analysis is $\Delta r = 0.0027 \pm 0.0019$ \cite{BKXi}.

From 2016 to 2018, we observed the CMB with \bicepthree\ at 95 GHz and \keck\ at 150, 220, 230, and 270 GHz (with most of the \textit{Keck} data in that time at 220/230 GHz).  
Although the \textit{Keck} 220 and 230 GHz bands are very similar, the spectral response of the receivers comprising these two bands have slightly different band centers, so they will be treated separately here. In future analyses they may be combined and presented as a single band, after verification that the impact of this combination on the multicomponent likelihood analysis is negligible.

In these proceedings, we present a high-level analysis of the far-field beam measurements taken on \bicepthree\ and \keck\ during this time, including progress on quantifying the \tp\ leakage in these three years of data.
We focus on the \bicepthree\ 95 GHz and \textit{Keck} 220 and 230 GHz bands.
In Section \ref{sec:optical}, we give a brief overview of the differences in optical design between \keck\ and \bicepthree.
We present far-field beam measurements in Section \ref{sec:ffbm}, with emphasis on the differential beam patterns before and after deprojection.
The preliminary results of the beam map simulations, including leakage $Q$ maps, are shown in Section \ref{sec:beamsim}.

%% file: sec_optical.tex
\section{OPTICAL DESIGN}
\label{sec:optical}

\bicepthree\ and \keck\ both feature compact, two-lens on-axis refracting telescopes, although there are some differences in their optical design.  
Overall, \bicepthree\ has a larger aperture (520 mm vs 264 mm for \textit{Keck}), faster optics ($f/1.7$ vs $f/2.4$), and a wider field of view ($27\deg$ vs $15\deg$).
In both cases, the lenses are cooled to 4 K during observations, and various IR filters at both 50 K and 4 K.

\begin{figure} [ht]
  \begin{center}
  \includegraphics[height=10cm]{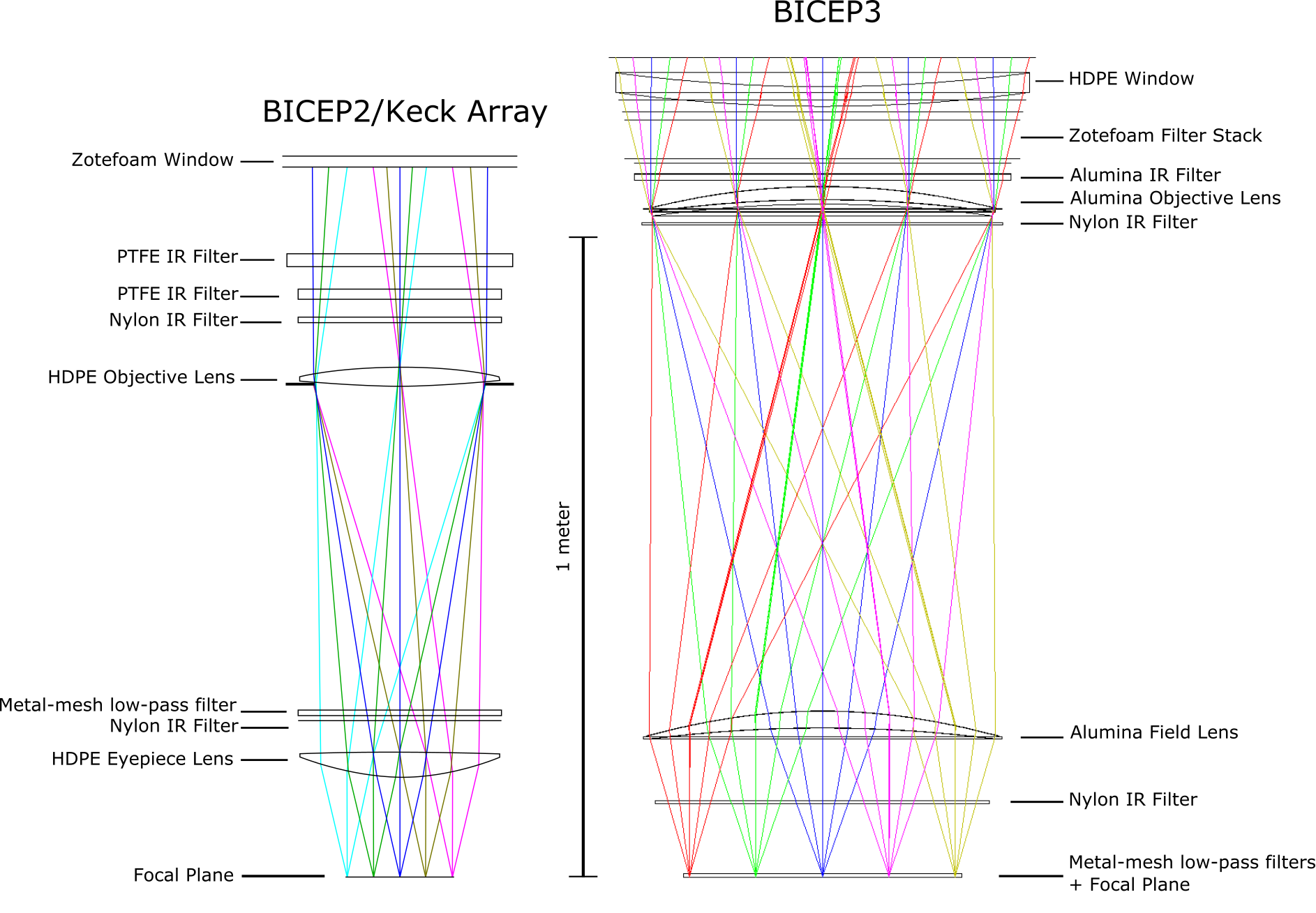}
  \end{center}
  \caption[example] 
   { \label{fig:optics} 
    Optical diagram for \keck\ (left) and \bicepthree\ (right).
    Individual optical elements are labeled and drawn roughly to scale.
    The Zotefoam filter stack replaced the stack of metal-mesh reflective filters in the 2016-2017 austral summer.
    The window was replaced with a \biceparray-compatible window (also made of HDPE) in the 2018-2019 austral summer.
    This figure is a slightly modified version of Figure 1 in Ref.~\citenum{spie:ksk}.}
\end{figure} 

Figure~\ref{fig:optics} shows optical diagrams comparing \bicepthree\ and \keck\ telescopes.  
At 50 K, \textit{Keck} has a series of polytetrafluoroethylene (PTFE) and nylon IR filters, while \bicepthree\ uses a high-density polyethylene foam filter stack.  
This foam filter stack replaced the metal-mesh IR-reflective filters originally installed in \bicepthree, as part of an upgrade in the 2016-2017 austral summer.  
This resulted in significantly reduced thermal loading on the 50 K stage \cite{spie:jhk}.
The lenses are kept at 4 K in both telescope designs, with \bicepthree\ using alumina ceramic and \textit{Keck} using high-density polyethylene (HDPE).  
Also at 4 K are nylon IR filters and metal-mesh low-pass filters to prevent high-frequency photons from coupling directly to the detectors.
The \bicepthree\ window is made of HDPE and is anti-reflection coated with expanded PTFE\footnote{Teadit 24GRD www.teadit.com}.  
In the 2018-2019 austral summer, this window was upgraded to the slightly larger window design of \biceparray, without changing any other optical components.
This allows future compatibility between \bicepthree\ and \biceparray\ windows and their associated hardware.

%% file: sec_ffbm.tex
\section{FAR FIELD BEAM MEASUREMENTS}
\label{sec:ffbm}

Every austral summer before starting CMB observations, we dedicate 1-2 months to measuring the far-field beam patterns of all our telescopes \textit{in situ}.
The measurement process is described in detail in the BK15 Beams Paper \cite{BKXi}; here we give a brief overview.
The small aperture design yields a far-field distance $2D^2/\lambda < 200$ m for all observing frequencies.
\bicepthree\ and \keck\ are housed in buildings at the South Pole that are $\sim$ 200 m apart, allowing us to observe a chopped thermal source from the opposite building to measure the beam response.
A large, flat aluminum mirror is used to redirect the beams over the groundshield to the thermal source on the opposite building.  
The mirror over \bicepthree\ intercepts all beams simultaneously, whereas the mirror over \textit{Keck} only intercepts the beams from, at most, two of the five receivers at a time.

After demodulating the raw beam map timestreams at the chop rate (usually 16 Hz), we bin into ``component" maps with $0.1\deg$ square pixels using an instrument-fixed coordinate system.  
The portions of the component maps where the main beam intersects the ground are masked and removed, to prevent possible contamination after demodulation.
Each component beam map is fit to a 2D elliptical Gaussian, allowing us to get estimates of beam width, pointing, and ellipticity for each detector.
Detailed measurements of the beam parameters and differential ($V - H$) beam parameters for \bicepthree\ and \keck\ between 2016 and 2019 can be found in Ref.~\citenum{ltd:tsg}.

In a given year of beam map measurements, one detector may have anywhere between $2-10$ component beam maps ($2-40$ for \bicepthree\, due to the better mirror coverage) after automatically cutting poor data using the fit Gaussian parameters.
These component maps, which are taken at various boresight rotation angles, are all coadded into per-detector composite beam maps, which are the highest-fidelity measurements of the main beam response we have for each detector.
These composite beam maps are used in the beam map simulations described in Section \ref{sec:beamsim}, and after being coadded over all detectors in a frequency band, are used to evaluate the beam window functions.
The \bicepthree\ and \keck\ beam window functions between 2016 and 2019 are also found in Ref.~\citenum{ltd:tsg}.

It is the difference in beam response between the $V$ polarization detector and $H$ polarization detector that leads to \tp\ leakage in the CMB data.
We use a technique called deprojection to filter out the leading-order terms of this leakage (originally defined in BK-III \cite{BKIII}).
To second order, the modes of a differential elliptical Gaussian couple to linear combinations of CMB $T$ and its first and second derivatives.  
Since the beam shapes are constant in time, we can construct leakage template maps corresponding to these difference modes, sample them using each detector pair's real trajectory data, regress each detector pair's signal timestream against its leakage template, then subtract the fitted template from the signal.  
Any power remaining in the difference beam response after deprojection contributes \tp\ leakage to the CMB data which is not filtered or removed by this technique.
Through the rest of this proceeding we refer to this post-deprojection difference power as the ``undeprojected residuals" or the ``residual beam."

\begin{figure} [t]
  \begin{center}
  \includegraphics[height=13cm]{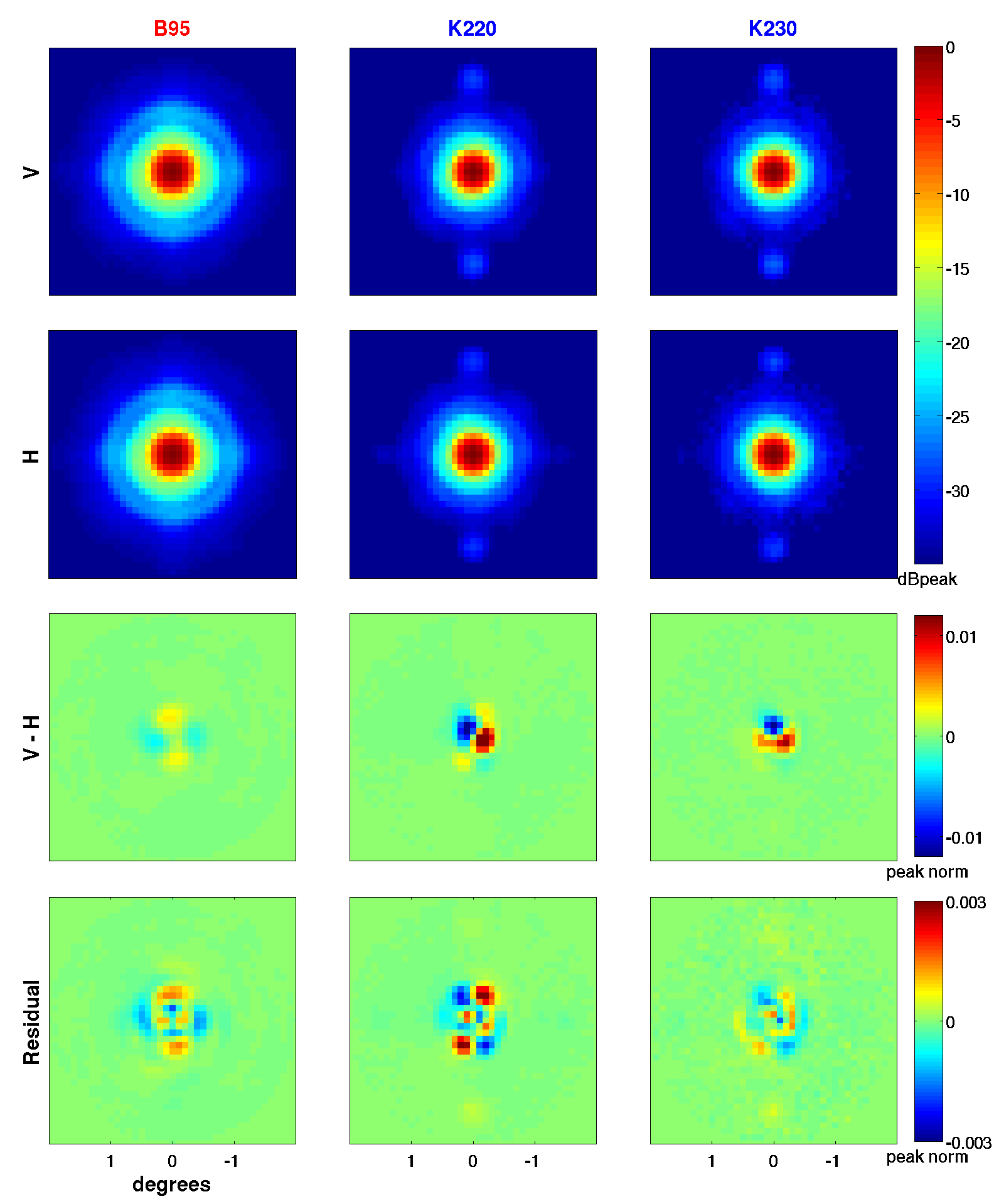}
  \end{center}
  \caption[example] 
   { \label{fig:beams1} 
    Plots of the $V$ and $H$ polarization beams, difference ($V - H$) beams, and undeprojected residuals averaged across all detectors in a band.
    The left column shows averages over \bicepthree, the middle column shows averages over \textit{Keck} 220 GHz, and the right column shows averages over \textit{Keck} 230 GHz.  
    Note the different color scale in the $V - H$ plots and the residual beam plots.
    The circular features at $\sim -25$ dB that are roughly 1 degree from the main beam are due to crosstalk in the time-domain readout system and have been previously characterized \cite{BKIII}.
    }
\end{figure}

Before running full beam map simulations, one can compare the shapes and magnitudes of the difference beams ($V - H$) and the undeprojected residuals to get a rough sense of leakage expected before and after deprojection.  
Figure~\ref{fig:beams1} shows the $V$, $H$, $V - H$, and residual beams averaged over all detectors in \bicepthree, \textit{Keck} 220 GHz, and \textit{Keck} 230 GHz.
In general, we see that differential pointing dominates the $V-H$ beam patterns for \textit{Keck} 220 GHz and \textit{Keck} 230 GHz, but less so for \bicepthree. 
This is consistent with previous beam parameter estimates that showed \bicepthree\ had much lower differential pointing than \textit{Keck} \cite{ltd:tsg}.
The average residual beam power in \textit{Keck} 220 and 230 GHz has a very similar pattern, with slightly higher amplitude at 220 GHz.  
This pattern is currently being investigated.
Although there is value in quantifying the relative average difference beam power between frequency bands, averaging over entire receivers may disguise complex variations or trends in these patterns across the detector array.
There is ongoing work to evaluate simple, per-detector metrics of \tp\ leakage derived directly from the beam maps, to quantify these possible trends and identify possible correlations with other optical measurements.

Another factor that determines the leakage contributing to the final CMB maps from the undeprojected residuals is the coaddition over observations using multiple boresight rotation angles.
Depending on the symmetry of a residual beam pattern and the amount of data at each angle contributing to the map, coaddition over many different angles may significantly reduce the total \tp\ leakage from a detector pair in a given map pixel.
\keck\ observes the CMB over eight boresight rotation angles each separated by $45\deg$, which allows cancellation of any residual beam patterns that are invariant under $90\deg$ or $180\deg$ rotations.
However \bicepthree\ only observes at four boresight angles (two separated by $45\deg$ and two that are their $180\deg$ compliments), due to the geometry of the cryocooler within the mount restricting boresight rotation.
This means cancellation of residual beams patterns invariant under $90\deg$ rotations cannot be achieved for \bicepthree. 
For a more complete discussion on the expected degree of cancellation of different leakage modes, see Ref.~\citenum{BKIII}.
%A beam component that is invariant under rotation, switches sign under $180\deg$ rotation, or switches sign under $90\deg$ rotation has monopolar, dipolar, or quadrupolar symmetry, respectively.
%As discussed in detail in previous publications, leakage from monopolar modes cancels when coadding over rotation angles separated by $90\deg$, and leakage from dipolar modes cancels when coadding over angles separated by $180\deg$.  
%No cancellation of quadrupolar or higher modes can be achieved by coadding over multiple boresight angles  \cite{BKIII}.

\begin{figure} [t]
  \begin{center}
  \includegraphics[height=13cm]{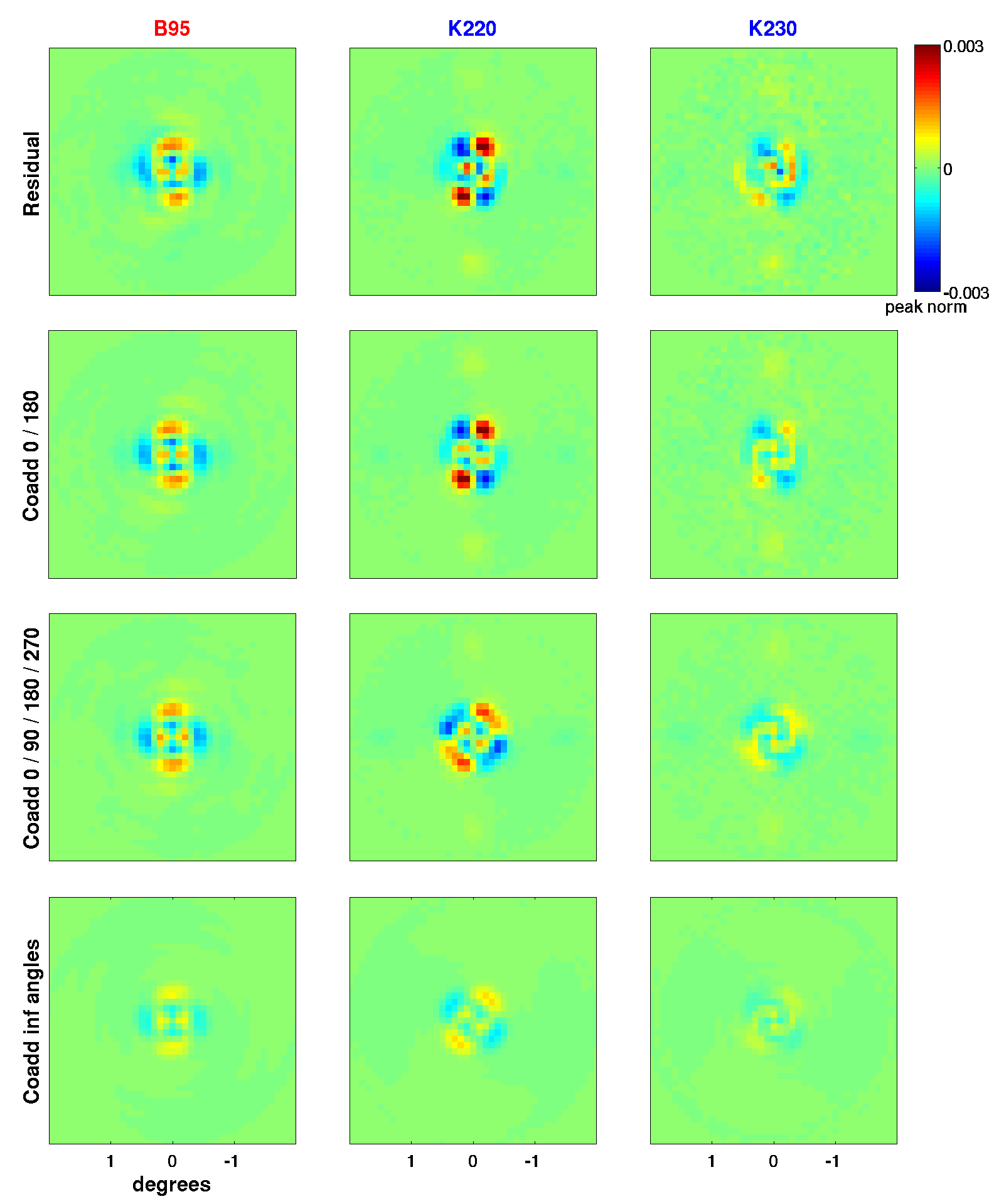}
  \end{center}
  \caption[example] 
   { \label{fig:beams2} 
    Plots of the undeprojected residuals averaged across all detectors in a band (\textit{1st row}).
    The remaining plots show the effective residual beam after coadding over two boresight rotation angles separated by $180 \deg$ (\textit{2nd row}), four boresight rotation angles separated by $90 \deg$ (\textit{3rd row}), and
    infinitely many boresight rotation angles (\textit{4th row})}
\end{figure} 

Figure~\ref{fig:beams2} shows the same band-averaged undeprojected residuals from the bottom row of Figure~\ref{fig:beams1}, now coadded over multiple boresight rotation angles to illustrate the effective total \tp\ leakage.
This coadding is done by averaging over the residual beam rotated to the specified boresight angles, weighted by $\cos{2 \theta}$, where $\theta$ is the boresight angle.
The similarity in the second row and third row for \bicepthree\ indicates that the penalty for using fewer rotation angles is small, on average.
This is expected for a quadrupole-like pattern as seen in the \bicepthree\ average residual beam -- a monopole-like pattern will cancel under $90 \deg$ rotations and a dipole-like pattern will cancel under $180 \deg$ rotations, but no such cancellation can be achieved with quadrupoles \cite{BKIII}.

%% file: sec_beamsim.tex
\section{BEAM MAP SIMULATIONS}
\label{sec:beamsim}

As CMB experiments increase detector count and sensitivity in future generations of telescopes, the ability to estimate and minimize \tp\ leakage is critical to our success at constraining cosmological parameters.  
We use ``beam map simulations" to quantify the expected amount of \tp\ leakage due to beam shape mismatch in our real CMB maps.
In this section, we present the latest leakage $Q$ maps resulting from these simulations, and discuss future efforts to estimate the resulting bias on $r$ in our upcoming BK18 (all data through the 2018 observing season) data set.

\begin{figure} [b]
  \subfloat{%
  \includegraphics[clip,width=\columnwidth]{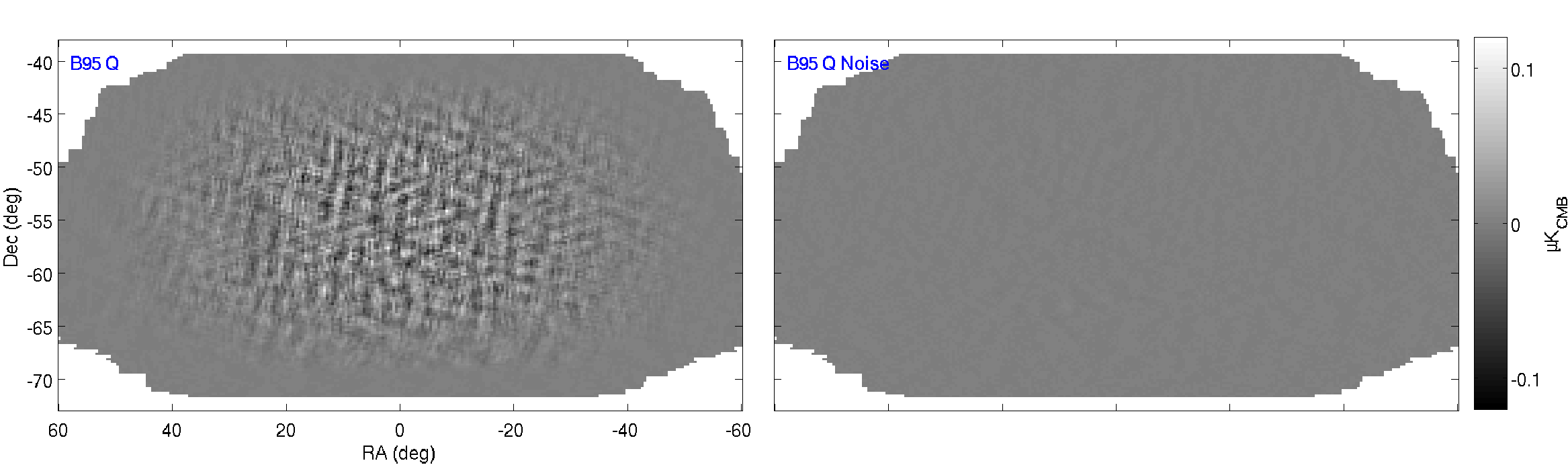}%
}

\subfloat{%
  \includegraphics[clip,width=\columnwidth]{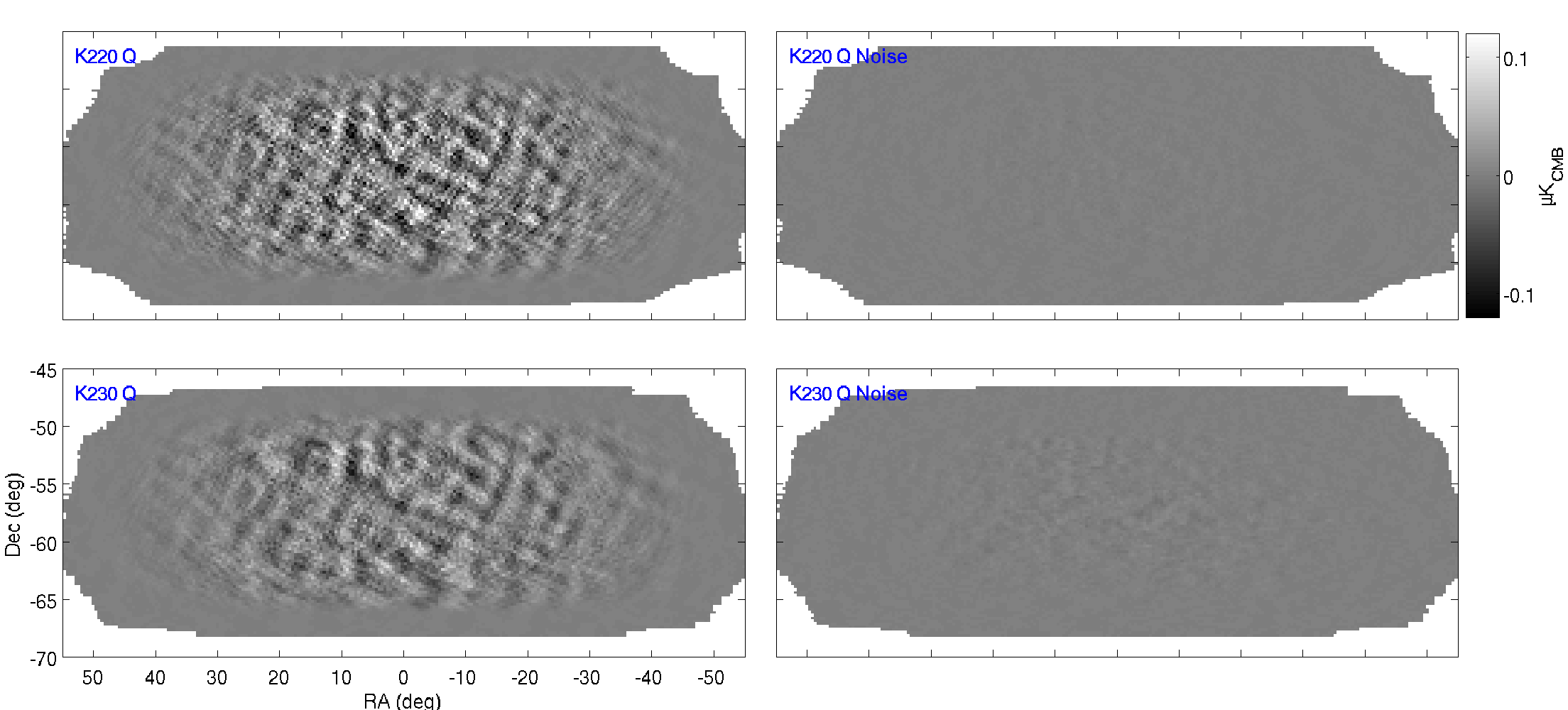}%
}
   \caption{ \label{fig:Qmap} 
    Apodized $Q$ maps of \tp\ leakage from beam map simulations coadded over the 2016, 2017, and 2018 observing seasons.  
    The left column shows the signal as predicted using composite beam maps, and the right column shows noise predicted by split beam maps.  \textit{Top:} \bicepthree\ 95 GHz, \textit{middle:} \textit{Keck} 220 GHz, \textit{bottom:} \textit{Keck} 230 GHz.  
    Note the scale of the RA and Dec axes are slightly different between the \textit{Keck} maps and the \bicepthree\ maps.
    All plots use the same colorscale.}
\end{figure}

The same analysis pipeline used for our real data and standard CMB simulations is also used for the beam map simulations.  
The input sky is the \textit{Planck} temperature map, with no polarization ($Q, U = 0$) and no added noise.
These input maps are convolved with the per-detector composite beam maps described in Section \ref{sec:ffbm} (truncated to a radius $< 2\deg$ from the main beam), then sampled using real detector pointing timestreams.
Just as with the standard CMB simulations, the data cuts and weights applied are taken from the real data.
The same deprojection method described in Section \ref{sec:ffbm} is also applied.
The timestreams are then binned into $Q/U$ maps, where any non-zero polarization signal must be due to mismatch in the composite beam shapes.

Although the input temperature maps in these simulations have no injected noise, the composite beam maps include both signal and noise from the beam map measurement.  
To estimate the amount of noise in the leakage $Q/U$ maps due to noise in the beam maps, we generate ``split" beam maps alongside the standard composite beam maps.
For a given detector, its split beam map is evaluated by randomly dividing all component maps (that pass automatic cuts) into two halves, and taking the difference.
These split maps are used in beam map simulations in the same way as the composite beam maps, and the resulting $Q,U$ leakage maps are treated as estimates of noise on the leakage estimates.

Figure~\ref{fig:Qmap} shows the resulting leakage $Q$ maps from beam map simulations coadded over the 2016, 2017, and 2018 seasons.  
The \textit{Keck} 220 GHz and 230 GHz leakage signal maps have very similar structure, which is somewhat expected due to the similarity of the average residual beams shown in Figure~\ref{fig:beams1} and Figure~\ref{fig:beams2}.
The noise in the beam maps (and therefore the leakage $Q$ noise map) is smaller for \bicepthree, mostly due to the higher number of component beam maps used to form the per-detector composites (due to the better coverage of the redirecting mirror).  
Although detailed power spectrum analysis of these maps is still in progress, preliminary analysis indicates that the amount of leakage seen in \bicepthree\ is smaller than any \textit{Keck} band in the relevant multipole range ($\ell \sim 50 - 200$), and that the leakage in \textit{Keck} 220 and 230 GHz seems consistent with previous published 220 GHz results \cite{BKXi}.

Future work for this \tp\ analysis includes calculating power spectra after using our matrix-based purification that reduces $E \rightarrow B$ leakage due to partial sky coverage and filtering.
We will then use the same quadratic estimator $\rho$ as defined in the BK15 Beams Paper \cite{BKXi} to evaluate the systematic contribution to the single-frequency $BB$ spectra.
Finally, we plan to take these maps (after coadding with previous observing seasons) and form cross spectra with the real BK18 $Q/U$ maps.

%% file: sec_conclusion.tex
\section{CONCLUSIONS}
\label{sec:conclusion}  

In these proceedings we have presented a preliminary, high-level analysis of the far-field beam response of the \bicepthree\ and \keck\ CMB polarimeters in the 2016, 2017, and 2018 observing seasons.
We show the frequency band-averaged $V$ polarization, $H$ polarization, $V - H$, and residual beam response for \bicepthree\ and the \textit{Keck} 220 and 230 GHz bands.
We then show leakage $Q$ maps from beam map simulations that are estimates of the \tp\ leakage expected in our real CMB maps.
As expected from comparisons of the residual beams, the amount of $Q$ leakage in \bicepthree\ between 2016 and 2018 is lower than that of \textit{Keck} 220 and 230 GHz (based on preliminary power spectrum analysis).
Leakage seen in 220 and 230 GHz is roughly consistent with previous \tp\ analysis of \textit{Keck} 220 GHz.